\documentclass[journal,twoside,web]{ieeecolor}

\usepackage{silence}
\usepackage{etoolbox}
\makeatletter
\@ifundefined{color@begingroup}%
{\let\color@begingroup\relax
\let\color@endgroup\relax}{}%
\def\fix@ieeecolor@hbox#1{%
\hbox{\color@begingroup#1\color@endgroup}}
\patchcmd\@makecaption{\hbox}{\fix@ieeecolor@hbox}{}{\FAILED}
\patchcmd\@makecaption{\hbox}{\fix@ieeecolor@hbox}{}{\FAILED}

\usepackage{tmi}
\usepackage{caption}
\usepackage{subcaption}
\usepackage{amsmath,amssymb,amsfonts}
\usepackage{algorithmic}
\usepackage{graphicx}
\usepackage{textcomp}
\usepackage{microtype}
\usepackage{cite}    
\usepackage{float} 
\usepackage{multirow}
\usepackage{bbding}
\usepackage{booktabs}
\usepackage{pifont}

\usepackage{amsmath}
\usepackage{amssymb}
\usepackage{bm}
\usepackage{cases}
\usepackage{mathrsfs}
\usepackage{booktabs}  
\usepackage{array}     
\usepackage{caption}   
\usepackage{graphicx}     
\usepackage{subcaption}   
\usepackage{siunitx} 
\usepackage{makecell} 
\usepackage{booktabs} 
\usepackage{makecell} 
\usepackage{siunitx}  
\usepackage{array}    
\usepackage{titlecaps}
\usepackage{cite}
\usepackage{float}
\usepackage{svg}
\usepackage{graphicx}
\usepackage{adjustbox}
\usepackage{pdfpages}
\usepackage{stfloats} 
\usepackage{multirow}
\usepackage{xcolor}
\usepackage{makecell}
\usepackage[percent]{overpic} 
\usepackage{colortbl}
\usepackage{balance}
\usepackage{xurl}

\makeatletter
\let\NAT@parse\undefined
\makeatother
\usepackage[colorlinks=true, allcolors=blue]{hyperref}
\definecolor{linkcolor}{RGB}{255,0,0}
\definecolor{urlcolor}{RGB}{255,105,180}
\definecolor{citecolor}{RGB}{0, 80, 200}
\definecolor{citecolor1}{RGB}{0,153,255}
\usepackage[svgnames,x11names]{xcolor}
\definecolor{darkpastelgreen}{rgb}{0.01, 0.75, 0.24}
\definecolor{darkpastelred}{RGB}{230, 50, 0}
\definecolor{green}{RGB}{10,179,38}
\newcommand{\cmark}{\ding{51}}%
\newcommand{\xmark}{\ding{55}}%
\usepackage{hyperref}
\hypersetup{colorlinks=true,linkcolor=linkcolor,urlcolor=urlcolor,citecolor=citecolor1}

\usepackage{relsize}

\def\BibTeX{{\rm B\kern-.05em{\sc i\kern-.025em b}\kern-.08em
    T\kern-.1667em\lower.7ex\hbox{E}\kern-.125emX}}
\begin{document}
\title{GLEAM: A Multimodal Imaging Dataset and HAMM for Glaucoma Classification }
\author{Jiao Wang, \IEEEmembership{Member, IEEE}, Chi Liu, Yiying Zhang, Hongchen Luo, Zhifen Guo, Ying Hu, Ke Xu, Jing Zhou, Hongyan Xu, Ruiting Zhou, Man Tang
\thanks{This work was supported by the National Natural Science Foundation of China (NSFC) under Grant 62502080.
\textit{(Jiao Wang and Chi Liu are co-first authors.) (Corresponding author: Jiao Wang.)}
 }%
 \thanks{Jiao Wang, Yiying Zhang, Hongchen Luo, Zhifen Guo, Hongyan Xu, Ruiting Zhou, Man Tang are with College of the Information Science and Engineering, Northeastern University, Shenyang 110819, China (e-mail: wangjiao@ise.neu.edu.cn; 2470818@stu.neu.edu.cn; luohongchen@ise.neu.edu.cn; 2210301@stu.neu.edu.cn; 2400757@stu.\\neu.edu.cn; 2400990@stu.neu.edu.cn; 2470932@stu.neu.edu.cn).}
  \thanks{Chi Liu, Ying Hu, Ke Xu, Jing Zhou are with the Department of Ophthalmology, Shenyang Fourth People’s
Hospital, Shenyang 110000, China (e-mail: huying.oculist@gmail.com; xuke830711@gmail.com). 
}
}
\maketitle

\begin{abstract}

Glaucoma is a leading cause of irreversible blindness worldwide, with asymptomatic early stages often delaying diagnosis and treatment. 
Early and accurate diagnosis requires integrating complementary information from multiple ocular imaging modalities. 
However, most existing studies rely on single- or dual-modality imaging, such as fundus and optical coherence tomography (OCT), for coarse binary classification, thereby restricting the exploitation of complementary information and hindering both early diagnosis and stage-specific treatment.  
To address these limitations, we propose glaucoma lesion evaluation and analysis with multimodal imaging (GLEAM), the first publicly available tri-modal glaucoma dataset comprising scanning laser ophthalmoscopy fundus images, circumpapillary OCT images, and visual field pattern deviation maps, annotated with four disease stages, enabling effective exploitation of multimodal complementary information and facilitating accurate diagnosis and treatment across disease stages. 
To effectively integrate cross-modal information, we propose hierarchical attentive masked modeling (HAMM) for multimodal glaucoma classification. Our framework employs hierarchical attentive encoders and light decoders to focus cross-modal representation learning on the encoder. 
The attention module, named multimodal-channel graph attention (MCGA), boosts glaucoma classification performance by emulating two key clinical reasoning steps: first, it uses a multi-head modality gating mechanism to replicate ophthalmologists’ confidence scoring of fundus, OCT, and VF modalities; then, MCGA leverages a relational graph attention network to cross-examine structural-functional consistencies of weighted modalities. 
The experiments on GLEAM demonstrate that tri-modal fusion significantly outperforms single-modal and dual-modal configurations. Moreover, our proposed HAMM achieves superior performance compared with state-of-the-art multimodal learning methods. 
The dataset and code are publicly available via \href{https://github.com/microewing/HAMM}{https://github.com/microewing/HAMM}.
\end{abstract}

\section{Introduction}

\IEEEPARstart{G}{laucoma} is an irreversible, chronic ophthalmic disease and a leading cause of blindness worldwide. 
Globally, glaucoma affects an estimated 70 million people\cite{Vos2016154570}, and epidemiological projections indicate that by 2040, more than 110 million individuals worldwide will suffer from varying degrees of visual impairment caused by this disease\cite{BARKANA2015e402024}. 
In clinical practice, diagnosis relies on a combination of imaging and functional assessments, including color fundus photography (CFP) for evaluating the optic nerve head, optical coherence tomography (OCT) for generating retinal nerve fiber layer (RNFL) thickness measurements to quantify structural thinning, and visual field (VF) testing for assessing functional deficits \cite{VCDR,RNFL,MD}. Accurate classification of glaucomatous visual field defects into different severity stages is clinically important, as it enables tailored treatment based on disease severity, facilitates more consistent follow-up, and provides a more reliable prognosis\cite{GlaucomaStagingReview}. However, traditional manual diagnostic approaches suffer from several challenges, including slow processing speed, inconsistent accuracy, and a heavy reliance on physician expertise, underscoring the urgent need for more efficient and reliable diagnostic solutions.

To address these challenges, computer-aided diagnostic (CAD) systems have gained increasing attention. Over the past decade, numerous publicly available glaucoma datasets have been introduced focusing on a single modality\cite{5626137ORIGA-Light,6867807Drishti-GS,ORLANDO2020101570Refuge,HDV1,9515188ACRIMA,10377279Harvard-GDP1000}. Correspondingly, single-modal approaches\cite{2018disc_aware_enssemble_network,2023GLIM-NET,2021CCT-NET,2022_work_flow,Glaucoformer,FJA-Net} have been developed for glaucoma diagnosis, achieving steady improvements in accuracy and interpretability.

Despite these advances, single-modal systems are limited by the heterogeneous manifestations of glaucoma. Early lesions can be imperceptible in certain modalities, and modality-specific noise or artifacts further compromise reliability \cite{artifacts,noise}. Structure–function discordance is also common, where optic nerve changes may precede functional deficits or vice versa \cite{Structure–function_relationship}. Clinical studies suggest that reliance on a single modality risks missing cases \cite{Bowd2008BayesianML,jcm11010216}. These limitations motivate multimodal integration, as demonstrated by the GAMMA challenge \cite{WU2023102938GAMMA} and subsequent multimodal fusion methods \cite{2023ELF,Wang_2023Mstnet,9761712Corolla,10388423GeCoM-NET,10.1007/978-3-031-73119-8_2ETSCL}, which outperform single-modal approaches. Collectively, these findings highlight the clinical relevance and technical promise of multimodal analysis for glaucoma.

However, existing multimodal glaucoma datasets \cite{RAJA2020105342AFIO,WU2023102938GAMMA,GRAPE,Harvard-GF10472539} remain limited in scale, modality diversity, and staging granularity, hindering the exploitation of complementary information, early diagnosis, and stage-specific treatment. 
This inadequacy contrasts with clinical practice, where ophthalmologists routinely integrate fundus imaging for optic disc morphology, OCT-derived RNFL thickness for axonal integrity, and VF testing as the gold standard for functional loss. Each modality provides unique but incomplete information, and their joint interpretation is critical for cross-validating suspicious findings, improving sensitivity to early or atypical cases, and mitigating errors caused by modality-specific artifacts or noise.

To overcome these limitations, we introduce \textbf{Glaucoma Lesion Evaluation and Analysis with Multimodal} Imaging (GLEAM) dataset, developed with Shenyang Fourth People’s Hospital. As shown in Tab. \ref{tab:Table 0}, GLEAM contains 1,200 samples, each with SLO fundus images, circumpapillary OCT images, and VF pattern deviation (PD) maps. All samples are annotated with a four-class glaucoma staging based on comprehensive clinical evaluation: non-glaucoma (NG), early glaucoma (EaG), intermediate glaucoma (InG), and advanced glaucoma (AdG).  All modalities are acquired within a short temporal window to ensure cross-modality consistency. With its scale, modality coverage, and fine-grained staging, GLEAM provides a robust resource for developing and benchmarking fusion-based algorithms for glaucoma detection and staging.

In addition, to address the clinically critical challenge of integrating structural and functional imaging information, we propose the \textbf{hierarchical attentive masked modeling (HAMM)} framework for multimodal glaucoma classification. Diverging from conventional transformer-based masked autoencoders (MAE), HAMM is built entirely on convolutional neural networks (CNNs), which facilitates the integration of skip connections, a novel multimodal-channel graph attention (MCGA) module, and light decoders. The design of the MCGA module is clinically inspired; it emulates the ophthalmologist's practice of (i) assigning adaptive confidence weights to fundus, OCT, and VF modalities based on their quality and reliability, and (ii) cross-examining structural-functional consistencies\cite{MEDEIROS2012814,Bayer2018}. Specifically, MCGA employs a multi-head modality gating mechanism and a relational graph attention network that captures inter-modal dependencies. This architecture emphasizes robust representation learning within the encoder, thereby yielding richer and more transferable representations for downstream glaucoma classification tasks.

In conclusion, the main contributions of this paper include the following three aspects:

\begin{itemize}

\item We developed GLEAM, a novel multimodal glaucoma dataset that for the first time integrates tri-modal imaging with finer-grained stage annotations, supporting more precise and reliable research on glaucoma-aided diagnosis.

\item We propose HAMM, which integrates a clinically inspired MCGA module and a designed MAE pretraining strategy to enable modality-aware fusion, achieving accurate and robust glaucoma classification.

\item We compare HAMM with existing state-of-the-art (SOTA) models on GLEAM, demonstrating both the advantages of the tri-modal GLEAM dataset and the effectiveness of the proposed method.

\end{itemize}

The remainder of this paper is organized as follows: We review related work in Section \ref{sec2}. In Section \ref{sec3}, we propose our dataset. Section \ref{sec4} presents the details of the proposed method. In Section \ref{sec5}, we report the experimental setup and results. Section \ref{sec6} provides a detailed discussion. Finally, we conclude this paper in Section \ref{sec7}.

\section{Related Work\label{sec2}}

\subsection{Glaucoma Datasets}

\begin{table*}[!t]
    \centering
  \renewcommand{\arraystretch}{1.}
  \renewcommand{\tabcolsep}{18 pt}
   \caption{Overview of major glaucoma datasets for classification. 
\textit{(NG: Non-glaucoma, EaG: Early Glaucoma, InG: Intermediate Glaucoma, AdG: Advanced Glaucoma , PG: Progressive Glaucoma, NPG: Non-progressive Glaucoma.})}
   \label{tab:Table 0}
 
  \begin{tabular}{r||ccc|cc}
    \hline
    \Xhline{2.\arrayrulewidth}
Dataset name&  Year&Quantity&Imaging modalities& Classification&Public\\
\hline
ORIGA\cite{5626137ORIGA-Light}& 2010& 650& Fundus& NG/G &{\color{darkpastelgreen} \cmark}\\
 Drishti-GS\cite{6867807Drishti-GS}& 2014& 101& Fundus& NG/G &{\color{darkpastelgreen} \cmark}\\
 Refuge\cite{ORLANDO2020101570Refuge}& 2018& 1200& Fundus& NG/G &{\color{darkpastelgreen} \cmark}\\
 HDV1\cite{HDV1}& 2018& 1542& Fundus& NG/EaG/PG &{\color{darkpastelgreen} \cmark}\\
 ACRIMA\cite{9515188ACRIMA}& 2019& 705& Fundus& NG/G &{\color{darkpastelgreen} \cmark}\\
 Harvard-GDP\cite{10377279Harvard-GDP1000}& 2023& 1000& RNFL& NG/G &{\color{darkpastelgreen} \cmark}\\
\hline
 D1\cite{2019CombinationOCT-bscan+fundus}& 2019& 78&  OCT \& Fundus& NG/G & {\color{darkpastelred} \xmark} \\
 AFIO\cite{RAJA2020105342AFIO}&  2020&50&OCT \& Fundus&  NG/G &  {\color{darkpastelgreen} \cmark} \\
 GAMMA\cite{WU2023102938GAMMA}&  2021&200& OCT \& Fundus& NG/EaG/PG & {\color{darkpastelgreen} \cmark}\\
 D2\cite{diagnostics12051100fundus+RNFL+comprehensive_examination}& 2022& 1006& Fundus \& RNFL& NG/G & {\color{darkpastelred} \xmark} \\
 D3\cite{KAMALIPOUR2023141VF+OCTA}& 2022& 1110& OCT \& OCTA& NPG/PG& {\color{darkpastelred} \xmark}\\
 GRAPE\cite{GRAPE}&  2023&144&RNFL \& VF&  NPG/PG&  {\color{darkpastelgreen} \cmark} \\
  Harvard-GF\cite{Harvard-GF10472539}& 2024&3300&OCT \& RNFL&  NG/G& {\color{darkpastelgreen} \cmark} \\
 D4\cite{HWANG2025100703jiazhou}& 2025& 706& Fundus \& RNFL 
 \& VF& NG/G& {\color{darkpastelred} \xmark} \\ 
 \hline 
 \textbf{GLEAM(Ours)}&  2025&1200& Fundus \& OCT 
\& VF&  NG/EaG/InG/AdG& {\color{darkpastelgreen} \cmark} \\\hline 
    \Xhline{2.\arrayrulewidth}
    \end{tabular}
  \end{table*}

Over the past decade, several single-modality glaucoma datasets were released for glaucoma classification. Fundus-based datasets \cite{5626137ORIGA-Light,6867807Drishti-GS,ORLANDO2020101570Refuge,HDV1,9515188ACRIMA} supported optic disc/cup segmentation and fundus-based glaucoma detection. OCT dataset\cite{10377279Harvard-GDP1000} mainly provided structural scans of the optic nerve head and RNFL for biomarker analysis. Specifically, Harvard Dataverse V1 (HDV1) dataset innovatively divided glaucoma into early and progressive stages, enabling the training of models with the capability to perform early-stage diagnosis, further advancing the potential for early intervention. While valuable, these single-modality datasets captured only partial aspects of glaucoma, motivating the development of multimodal datasets. 
The first small-scale multimodal glaucoma dataset, AFIO \cite{RAJA2020105342AFIO}, was released in 2020 and contained only 50 samples, each comprising a fundus and an OCT image, labeled as either non-glaucoma or glaucoma. In 2021, the GAMMA dataset \cite{WU2023102938GAMMA} was published with 200 samples and OCT/fundus modalities, which first stratified glaucoma into early and progressive stages in the multimodal domain, facilitating early diagnosis. 
The GRAPE dataset\cite{GRAPE}, published in 2023, addressed progression prediction by tracking patients over an average of 2.51 years. It classified 144 patients into progressive or non-progressive glaucoma based on longitudinal follow-up. Despite its longitudinal design, the cohort remained small. In 2024, the Harvard-GF dataset\cite{Harvard-GF10472539} expanded to 3,300 samples. It utilized OCT scans and derived RNFL maps for analysis. While larger, this dataset relies on two modalities derived from the same imaging source (OCT), limiting the upper limit of accuracy for multimodal classification. 
Other private datasets\cite{2019CombinationOCT-bscan+fundus,diagnostics12051100fundus+RNFL+comprehensive_examination,KAMALIPOUR2023141VF+OCTA,HWANG2025100703jiazhou} (denoted as D1, D2, D3, and D4) exhibit similarly limited sample sizes and remain inaccessible to external researchers. More details are provided in Tab. \ref{tab:Table 0}.

Existing datasets either suffer from small sample sizes, lack of accessibility, limited modality diversity, or staging granularity, which limit their utility for robust multimodal analysis. To address these gaps, we propose GLEAM, a novel public dataset comprising three image modalities and 1,200 clinical samples. This dataset introduces four categories: NG, EaG, InG, and AdG, representing the first multimodal glaucoma dataset to simultaneously offer the most image modalities and granular staging classification.

\subsection{Glaucoma Classification Methods}

Early glaucoma classification methods relied on a single modality, typically Fundus, OCT, or VF to train diagnostic models, 
achieving steady improvements in accuracy and interpretability through various architectures and longitudinal modeling\cite{2018disc_aware_enssemble_network,2023GLIM-NET,2021CCT-NET,2022_work_flow}. Some methods\cite{Glaucoformer,FJA-Net} use attention mechanisms to enhance glaucoma staging, achieving early diagnosis within a single modality. More recently, the focus has shifted towards multimodal fusion techniques\cite{Wang_2023Mstnet,2023Adongda东北大学,10388423GeCoM-NET,2023ELF,HWANG2025100703jiazhou,10030757MHCA,10944200DRIFA-Net}, where complementary information from different imaging sources is combined to achieve more accurate and robust glaucoma staging. Among these multimodal methods, attention-based techniques have emerged as a powerful approach to fuse information from different modalities. Attention mechanisms are particularly useful for capturing spatial and channel dependencies across modalities, enabling more accurate fusion and improving performance on complex tasks. MHCA\cite{10030757MHCA} applies joint multi-head spatial-channel attention through 2D convolution and deconvolution operations with multiple kernel sizes, enhancing detail capture in low-resolution environments. However, this method incurs a high parameter count due to the repeated downsampling and upsampling operations. DRIFA-Net\cite{10944200DRIFA-Net} incorporates robust residual attention and multimodal fusion attention in each convolution block, achieving hierarchical fusion, but at the cost of significantly higher parameters and computational time. To address these limitations, we propose the MCGA module, which integrates multimodal fusion more efficiently by incorporating an MCGA after each downsampling block in the CNN architecture. This design enables more effective hierarchical fusion. MCGA employs a multi-head modality-gating mechanism to assign adaptive confidence weights, and a relational graph attention network to capture inter-modal dependencies.

Furthermore, several studies\cite{NEURIPS2020_d89a66c7SCL,9761712Corolla,10.1007/978-3-031-73119-8_2ETSCL,He_2022_CVPRMAE,EyeCLIP,MultiMAE,UrFound,tian2023designingSparK} employ pre-training methodologies to enhance feature representation learning, subsequently fine-tuning models for classification tasks to improve glaucoma diagnostic accuracy. A prominent approach, supervised contrastive learning (SCL) \cite{NEURIPS2020_d89a66c7SCL,9761712Corolla,10.1007/978-3-031-73119-8_2ETSCL}, pulls embeddings from the same class closer together while pushing embeddings from different classes apart in the latent space, thereby maximizing inter-class discriminability. But unlike the classiﬁcation with natural images, where the images with different labels are likely visually distinguishable, there is only a little difference in ophthalmology pictures with different labels\cite{10377279Harvard-GDP1000}. This results in little to no role for the SCL method in the glaucoma classification task. Self-supervised learning (SSL), forcing models to learn contextual semantics without labels, offers an alternative pathway. MAE\cite{He_2022_CVPRMAE,EyeCLIP,MultiMAE,UrFound,tian2023designingSparK} is a representative SSL approach that learns contextual semantics through randomly masking image regions. Traditional MAE\cite{He_2022_CVPRMAE,EyeCLIP,MultiMAE,UrFound} is restricted to Transformer-based architectures, which lack visual inductive biases and rely heavily on large-scale annotated datasets. This makes MAE prone to overfitting on small-sample medical datasets, where labeled data is limited and costly to obtain. Additionally, MAE suffers from high computational complexity and struggles to incorporate hierarchical feature learning, limiting its efficiency and performance. SparK \cite{tian2023designingSparK} extended MAE to CNNs through a convolutional encoder-decoder, enabling hierarchical feature learning and achieving superior performance on downstream tasks. However, to the best of our knowledge, CNN-based MAE has not yet been applied to multimodal tasks. To address these limitations, we propose HAMM, which leverages CNN-based masked modeling with skip connections and the clinically inspired MCGA module, enabling adaptive weighting and cross-checking of multimodal data for more reliable and transferable representations.

\section{Glaucoma Dataset}
\label{sec3}

In this section, we introduce the GLEAM dataset. It comprises SLO fundus images, circumpapillary OCT images, and VF PD maps collected from a standardized clinical cohort with procedures for data acquisition, quality control, and image processing.

This study is approved by the Institutional Review Board (IRB) of Shenyang Fourth People’s Hospital (Approval Number: 2025-ZFKY-021; Approval Date: 2025.09.08) and adheres to the Declaration of Helsinki. The IRB waives the requirement for informed consent due to its retrospective design. All unique patient identifiers are permanently deleted from the dataset to ensure the privacy and confidentiality of patients. 
GLEAM serves as a benchmark for developing multimodal staging models to improve glaucoma diagnosis and is publicly available under a CC BY-NC-ND 4.0 license.  

\subsection{Data Collection and Quality Control}
\begin{figure}
    \centering
    \includegraphics[width=1\linewidth]{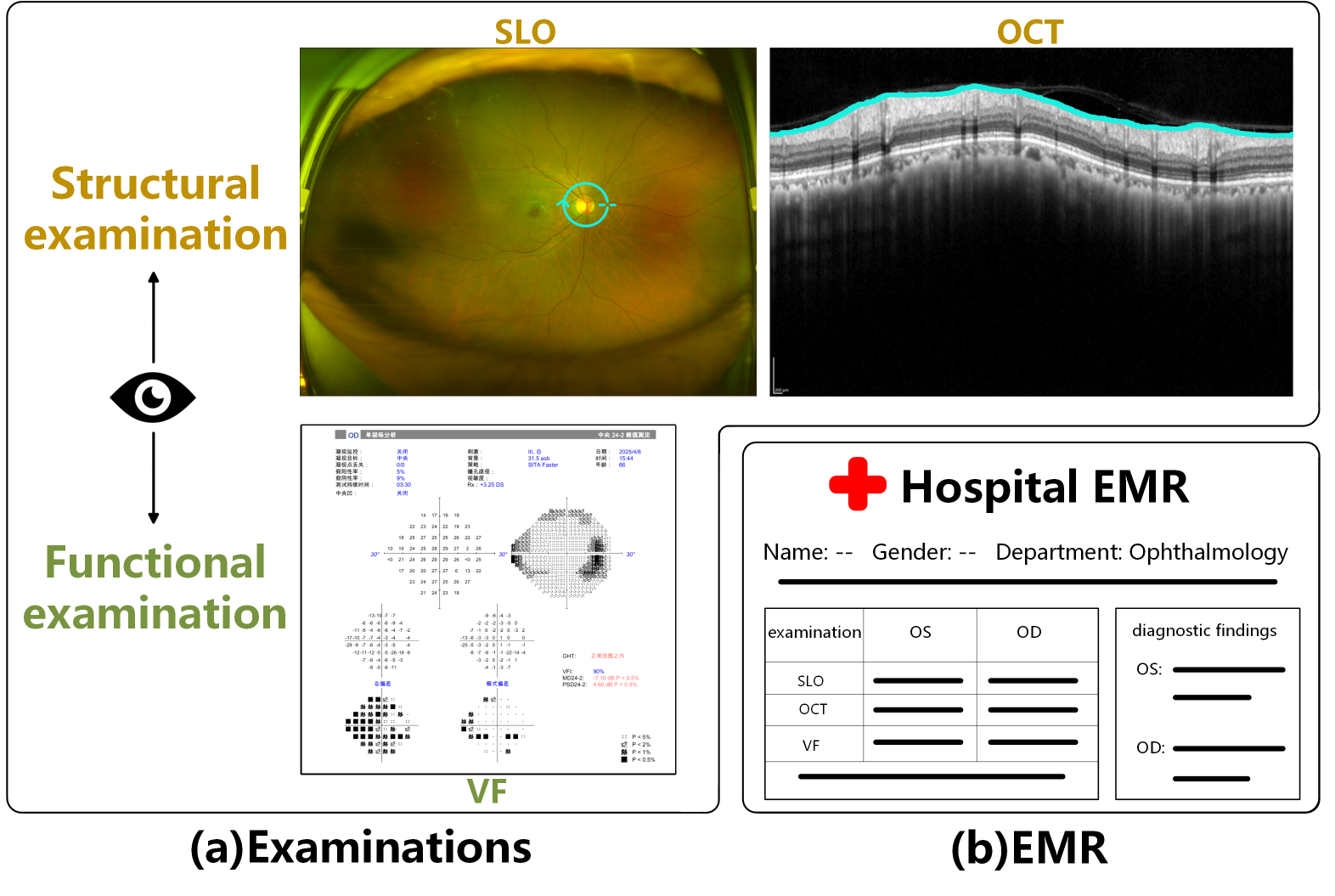}
    \caption{The detail of ophthalmic data, including (a) SLO, OCT, and VF examinations (b) EMR documented by ophthalmologists.}
    \label{fig 1}
\end{figure}
As shown in Fig. \ref{fig 1}(a) and Fig. \ref{fig 1}(b), we retrospectively collect multimodal ophthalmic data from Shenyang Fourth People’s Hospital, China, including:
\begin{itemize}
    \item \textbf{S}canning \textbf{L}aser \textbf{O}phthalmoscopy \textbf{(SLO)} Images: Acquired using the Optos ultra-widefield SLO device, providing high-resolution, wide-angle fundus imaging.
    \item Circumpapillary \textbf{O}ptical \textbf{C}oherence \textbf{T}omography \textbf{(OCT)} Images: Captured with the Heidelberg Spectralis OCT system, enabling detailed 
retinal layer analysis.
    \item \textbf{V}isual \textbf{F}ield \textbf{(VF)} Reports: Generated using Zeiss perimeter devices, with key metrics such as mean deviation (MD) and pattern standard deviation (PSD) recorded.
    \item \textbf{E}lectronic \textbf{M}edical \textbf{R}ecords \textbf{(EMR)}: Documented by ophthalmologists, including annotations of the patient 
details and ophthalmologists' diagnostic findings.
\end{itemize}

To establish a robust dataset, rigorous quality control focused on diagnostic integrity and label consistency is implemented. We included only samples where the SLO and circumpapillary OCT images are clear, and both the false negative and false positive rates of VF are less than 33\%. The labeling process consists of two stages: initial labeling and expert review. 

In the initial labeling stage, four trained annotators evaluate comprehensive diagnostic findings for each potential participant, categorizing them into one of three groups: normal, presence of glaucoma-related findings, or other ophthalmopathies. To minimize confounding effects and isolate manifestations specific to glaucoma pathology, all samples designated as other ophthalmopathies are systematically excluded prior. Subsequently, retained samples are assigned to one of four categories: NG, EaG, InG, and AdG. This assignment integrates available diagnostic findings and quantitative metrics. Explicit diagnostic findings of disease stage within the EMR form the primary basis for label assignment. For samples lacking explicit stage specification, categorization relies on the MD value from VF reports using clinical stratification thresholds: early-stage (MD \(>\) -6 dB), intermediate-stage (-12 dB \(\leqslant\) MD \(\leqslant\) -6 dB), and advanced-stage (MD \(<\) -12 dB)
\cite{stages1}.

Following initial labeling, three senior ophthalmologists independently review all multimodal images and assign stage labels. This review 
(i) excludes samples with poor quality (e.g., defocus, low illumination, misalignment, or severe artifacts), and (ii) verifies label consistency with both image findings and clinical context. Any discrepancies are reconciled through expert consensus, thereby improving diagnostic accuracy and resolving ambiguities. To comprehensively assess the labeling quality, two types of consistency evaluations are performed. For inter-grader agreement, the pairwise consistency of the first-round independent annotations from the three experts is calculated to reflect the consensus level among different clinicians. For intra-grader agreement, the three experts are asked to perform a second-round blind annotation on all retained samples after an interval of one month; it should be emphasized that this second-round annotation is conducted without access to their previous labeling records, and is used solely for consistency verification rather than label modification—the formal labels used in all subsequent experiments were determined by the first-round annotations after expert consensus. 
The quantitative results of both inter- and intra-grader consistency are summarized in Tab. \ref{tab:Table experts kappa}, where the quadratic weighted Cohen’s kappa coefficient is adopted to account for the ordinal nature of glaucoma staging. 

\begin{table}[!t]
    \centering
    \renewcommand{\arraystretch}{1.2}
    \renewcommand{\tabcolsep}{8pt}
    \caption{Inter- and Intra-grader Agreement via Quadratic Weighted Cohen’s Kappa (\%)}
    \label{tab:Table experts kappa}
    
    \begin{tabular}{r|c|c|c}
        \hline
        \Xhline{2.\arrayrulewidth}
        Agreement Type & Expert 1 & Expert 2 & Expert 3 \\ 
        \hline
        Intra-grader Agreement & 98.25 & 97.68 & 97.42 \\
        \hline
        Expert 1 vs. Expert 2 & 95.81 & - & - \\
        Expert 1 vs. Expert 3 & 96.37 & - & - \\
        Expert 2 vs. Expert 3 & - & 95.59 & - \\
        \hline
        \Xhline{2.\arrayrulewidth}
    \end{tabular}
\end{table}

\subsection{Image Processing for Dataset}

The processing pipeline mainly focuses on SLO images and VF reports, aiming to remove irrelevant information in the images. For SLO images, the optic disc centroid is manually adjusted to align with the image center across the dataset. A uniform region of interest (ROI) of 450×450 pixels is extracted, centered on the optic disc through a standardized cropping procedure. This ensures that the optic disc–cup region, the key area evaluated by ophthalmologists for glaucoma diagnosis \cite{8756196LAG}, is consistently retained across samples, while peripheral retinal areas prone to distortion and vignetting in ultra-widefield imaging are removed. 

For VF reports, an automated clipping procedure isolates the clinically critical Pattern Deviation map based on predefined coordinate boundaries relative to the report layout. The extracted 920×920-pixel image retains numerical axes and labels. To prepare a clean image suitable for pixel-based analysis, these axes are removed by overwriting rectangular regions containing axis lines and tick marks with a constant white mask, retaining only the grayscale sensitivity value grid.

\subsection{Dataset Character}

\begin{figure*}
    \centering
    \includegraphics[width=1\linewidth]{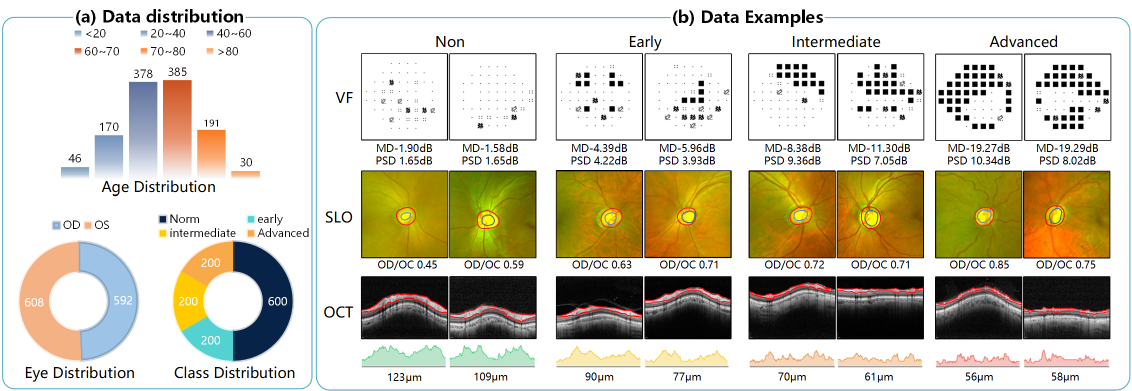}
    \caption{GLEAM dataset character, including (a) distributions of GLEAM, and (b) examples of GLEAM. We add lines for splitting the optic disc/optic cup and the thickness of the RNFL to more visually show the gaps in the images of different progressions of glaucoma. }
    \label{fig 3}
\end{figure*}

The final dataset comprises 1,200 multimodal samples, each integrating VF PD maps, SLO fundus images, and circumpapillary OCT images. As shown in Fig. \ref{fig 3}(a), GLEAM dataset includes 600 NG, 200 EaG, 200 InG, and 200 AdG samples, corresponding to 841 Chinese patients (56\% female). Patient ages ranged from 8 to 90 years (mean 55.4 ± 16.7 years). Of these,  482 patients provided data from both eyes, 249 patients provided oculus sinister (OS) data, and 233 patients provided oculus dexter (OD) data. This constitutes the first publicly available glaucoma dataset that simultaneously includes tri-modal imaging. The finer-grained four-classification system can also support more accurate diagnoses and treatments.

Fig. \ref{fig 3}(b) illustrates characteristic differences across glaucoma stages for each imaging modality.  VF PD maps exhibit increasing numbers and spatial extent of dark patches, corresponding to progressive retinal ganglion cell dysfunction and declining visual sensitivity. Concurrently, SLO fundus images reveal progressive enlargement of the optic disc-to-cup (OD/OC) ratio, resulting from optic cup excavation due to axonal loss in the optic nerve head. Circumpapillary OCT images demonstrate progressive thinning, directly correlating with neural damage severity. Obviously, the visual field report is the easiest to distinguish among the EaG, InG, and AdG modalities, but its interpretation is subjective and depends on the patient's degree of cooperation. 
Therefore, integrating complementary information from multiple modalities enables a more accurate assessment of glaucoma progression.

\section{Proposed Method\label{sec4}}

\begin{figure*}[!t]
    \centering
    \includegraphics[width=1\linewidth]{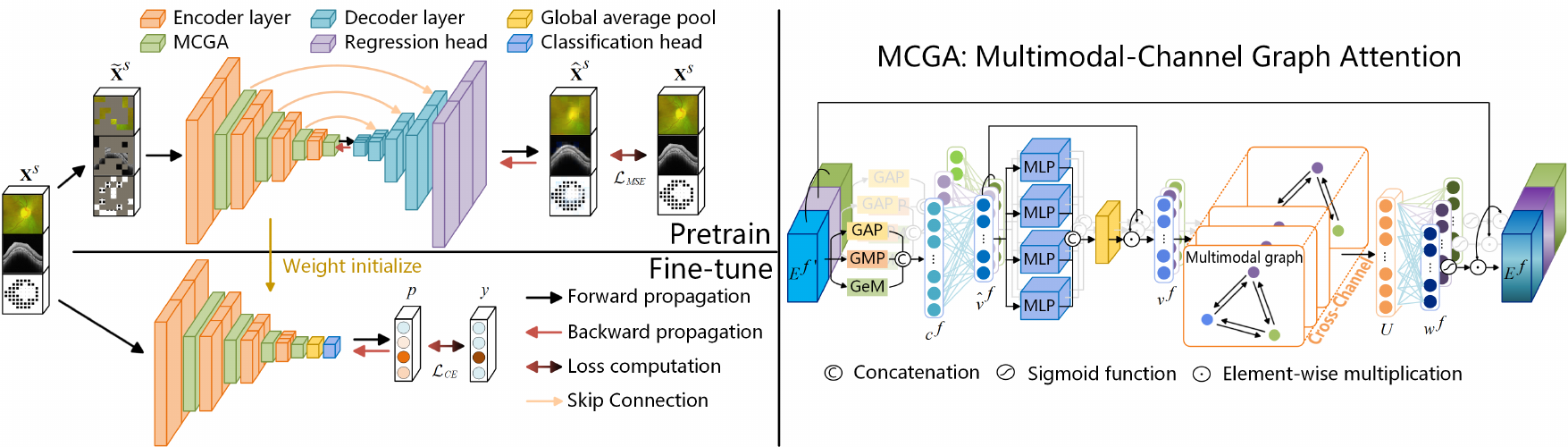}
    \caption{The architecture of HAMM. Stage one is the masked autoencoder pretraining task to train a better feature representation, and stage two is the classification task to train the final classification model. The MCGA module is designed to model the multimodal information between different images when pretraining and classification. For clarity, token notations (e.g $c^f$, $\hat v^f$, $v^f$) are shown for a single modality, though the same applies to all.}
    \label{fig 4}
\end{figure*}

Fig. \ref{fig 4} shows the proposed HAMM. The inputs of HAMM are fundus images, circumpapillary OCT images, VF PD maps, and the corresponding labels 
represented as \(\mathcal{D} = \{ (\mathbf{X}_i, y_i)\}_{i=1}^N\), where \(\mathbf{X}_i=\{ X^{f}_{i}, X^{o}_{i}, X^{v}_{i}\}\) denotes the three image modalities of the \(i\)-th sample, \(y_i\) is its corresponding label, and \(N\) is the total number of samples. 
The target tasks are specifically addressed with \(y_i \in \{0,1,2,3\}\), corresponding to NG, EaG, InG, and AdG. During training, the inputs \(\mathbf{X}_i\) are used to generate enhanced multimodal representations \(\mathbf{X}_i^S=\{ S^{f}_{i}, S^{o}_{i}, S^{v}_{i}\}\). For validation and testing, \(\mathbf{X}_i^S=\mathbf{X}_i=\{ X^{f}_{i}, X^{o}_{i}, X^{v}_{i}\}\). 
The goal of our proposed method is to optimize parameters \(\mathbf{\theta}\)  to learn a mapping function: \( \mathcal{F}_\mathbf{\theta}: \mathbf{X}_i \rightarrow p(y_i|\mathbf{X}_i)\), where \( p(y_i|\mathbf{X}_i) \) is the class probability distribution. We utilize three ResNet-50 \cite{2016ResNet} as encoders represented as \(\psi^{f}(\cdot),\psi^{o}(\cdot),\psi^{v}(\cdot)\). These encoders integrate MCGA modules at every layer to enhance feature representation through cross-modality dependency learning. The method then operates in two distinct stages: At stage one, we use a designed masked autoencoder architecture, enabling feature representation learning through reconstruction tasks. The masked inputs of three modalities represent as \(\mathbf{\tilde X}_i^S\) are encoded through \(\psi^{f}(\cdot),\psi^{o}(\cdot),\psi^{v}(\cdot)\) to extract feature maps \(\mathbf{\tilde E}_{1(i)}\). The feature maps then through our designed decoders \(\phi^{f}(\cdot),\phi^{o}(\cdot),\phi^{v}(\cdot)\) to reconstruct the feature maps into \(\mathbf{D}_{1(i)}\), minimized by MSE loss to approximate \(\mathbf{X}_i^S\). At stage two, we employ the pre-trained backbone from stage one to extract feature maps \(\mathbf{E}_{1(i)}\) from inputs \(\mathbf{X}_i^S\) and then classify by a classification head. Generally, the framework comprises three core components: (1) multimodal feature extraction with MCGA, (2) masked autoencoder pretraining for multimodal representation learning, and (3) fine-tuning for multimodal glaucoma classification. 

\subsection{Multimodal Feature Extraction with MCGA}

We employ three parallel ResNet-50 networks as modality-specific encoders to extract hierarchical features from fundus images, OCT images, and VF PD maps. Each encoder consists of four sequential layers, progressively transforming input modalities into semantically enriched feature representations through dimensional reduction and receptive field expansion. For layer $i$, the modality-specific feature maps are computed as follows:
\begin{equation}
\small
E_i^{k'} = \psi_i^{k}(E_{i-1}^k), \quad
\mathbf{E}_i =  \mathbf{ \xi}_i(\mathbf{E}_i^{'}),
\end{equation}
where $i=1,2,3,4$ indexes the layer, $k \in \{f,o,v\}$ denotes the modality, 
$\psi_i^{k}(\cdot)$ is the $i$-th layer of the $k$-th modality encoder, 
$\mathbf{E}_i^{'} = \{E_i^{f'},E_i^{o'},E_i^{v'}\}$ is the collection of modality-specific features before attention refinement, 
$\mathbf{\xi}_i(\cdot)$ is the MCGA module for multimodal-channel attention refinement, 
and $\mathbf{E}_i = \{E_i^{f}, E_i^{o}, E_i^{v}\}$ is the refined collection of modality-specific features after attention. Specifically, $\mathbf{E}_0=\mathbf{X}^S$ denotes the collection of input images. 
The final outputs of the encoders are $\{E_4^{f}, E_4^{o}, E_4^{v}\}$.

For MCGA module, to capture complementary channel-wise information, each modality feature map $E^{k'}$ is summarized via global average pooling (GAP), global max pooling (GMP), and generalized mean (GeM) pooling as follows:
\begin{equation}
\small
c^k = [GAP(E^{k'})\,\|\, GMP(E^{k'})\,\|\,GeM(E^{k'})],
\end{equation}
where $\|$ denotes vector concatenation. The concatenated pooled vector $c^k$ is projected through a fully connected layer with Sigmoid activation to generate the modality embedding as $v^k$, ensuring that the modality features are aligned in a way that facilitates the proper functioning of the subsequent graph convolutional attention mechanism. 

Due to imaging artifacts, occlusions, or differences in modality-specific characteristics, each modality may exhibit varying degrees of uncertainty or quality. To account for the uncertainty inherent across multiple modalities, we introduce a multi-head, learnable gating mechanism that simulates the process of multiple ophthalmic experts assessing the reliability of each modality.  By applying this gating mechanism, we aim to model the confidence or reliability of each modality's representation and incorporate this uncertainty into the attention mechanism, thereby improving the model's ability to selectively focus on the most informative features from each modality. The multi-head learnable gating mechanism is applied as follows:

\begin{equation}
\small
\hat v^k = v^k \odot \frac{1}{H} \sum_{h=1}^{H} g^{(h)}(v^k),
\end{equation}
where $H$ is the number of heads, $g^{(h)}(\cdot)$ is the $h$-th gating function, and $\odot$ denotes element-wise multiplication.

The gated embeddings from all modalities are stacked into a matrix $V = [\hat v^f, \hat v^o, \hat v^v] \in \mathbb{R}^{3 \times C}$. We then adopt a relational graph attention mechanism to model the complex inter-modal dependencies inherent. Our graph attention formula incorporates an additional relation embedding term that supplements relation-specific information between different modalities, enabling the model to better capture the distinct correlation patterns among fundus, OCT, and VF modalities. By treating each modality as a node and their correlations as edges, graph convolution effectively captures these pairwise relationships, while the introduced relation embedding further enhances the model’s ability to distinguish and model the unique inter-modal associations critical for glaucoma diagnosis. For each edge $(i,j)$ with relation type $r_{ij}$ and attention head $h$, the compatibility score is computed as:

\begin{equation}
\small
e_{ij}^{(h)} = \text{LeakyReLU} \Big( (a^{(h)})^\top 
\big[W^{(h)} \hat v^{i} \,\|\, (W^{(h)} \hat v^{j} + R_{r_{ij}}^{(h)}) \big] \Big),
\end{equation}
where $W^{(h)} \in \mathbb{R}^{C \times C}$ is the per-head projection matrix, $R_{r_{ij}}^{(h)} \in \mathbb{R}^{C}$ is the embedding of relation type $r_{ij}$, $a^{(h)} \in \mathbb{R}^{2C}$ is the learnable attention vector. To obtain a proper weighting distribution, these raw scores are normalized across the neighbors of node  $i$ as follows:

\begin{equation}
\small
\alpha_{ij}^{(h)} = \frac{\exp(e_{ij}^{(h)})}{\sum_{k \in \mathcal{N}(i)} \exp(e_{ik}^{(h)})},
\end{equation}
where $\mathcal{N}(i)$ denotes the set of neighbors of node $i$.  The contextualized representation of node $i$ is then aggregated from its neighbors across all heads as follows:  
\begin{equation}
\small
u^k = \frac{1}{H} \sum_{h=1}^{H} \sum_{j \in \mathcal{N}(i)} 
\alpha_{ij}^{(h)} \, \big( W^{(h)} \hat v^{j} + R_{r_{ij}}^{(h)} \big).
\end{equation}

The updated embeddings from all modalities are stacked into a matrix $U = [u^f, u^o, u^v] \in \mathbb{R}^{3 \times C}$, which is subsequently passed through a fully connected layer with Sigmoid activation to produce the final channel attention weights $\{w^f, w^o, w^v\}$. 

Finally, the attention weights are applied to the original feature collection $\mathbf{E}'$ to produce the refined feature:

\begin{equation}
\small
E^{k} = E^{k'} \odot w^k,
\end{equation}
where $\odot$ denotes element-wise multiplication with broadcasting along the spatial dimensions.

\subsection{Masked AutoEncoder Pretraining for Multimodal Representation Learning}
In ophthalmic imaging, data can often be incomplete or compromised due to issues like ghosting, blurriness, and occlusions caused by anatomical structures. These are common across modalities such as fundus and OCT, making it difficult to capture high-quality images across all areas. To address these challenges, we incorporate a Masked AutoEncoder mechanism during the initial training phase. The MAE is particularly well-suited for ophthalmic images, which explicitly simulate missing or corrupted data by masking parts of the input image and training the model to reconstruct the occluded regions. This process encourages the model to learn robust, context-aware features that are less dependent on any single image region, thereby improving its ability to generalize to real-world scenarios where missing information or distortions are common. For each modality \(k\) in the multimodal input \(X^s\), we generate masked inputs using a structured masking approach:
\begin{equation}
\small
\tilde s^k = s^k \odot M_k,
\end{equation}
where \(M_k\) is a binary mask, \(||M_k||=\rho \times area(x_{ki})\), and \(\odot\) denotes element-wise multiplication. The masked inputs \(\mathbf{\tilde X}^S=\{ \tilde s^f, \tilde s^o,\tilde s^v\}\) are then fed into the encoder network to generate encoded feature maps \(\{\tilde E_4^{f},\tilde E_4^{o},\tilde E_4^{v}\}\). By dynamically fusing the information across modalities in the encoder network, MCGA enables the model to focus on the most informative regions, compensating for missing or degraded parts in any single modality.

Subsequently, the encoded feature maps are passed through decoders $\phi^{k}_i(\cdot)$ to reconstruct the original images. We adopt lightweight separate decoders for each modality instead of a single decoder that handles multimodal reconstruction simultaneously. This design is motivated by the fact that a single decoder would inevitably involve fusion of different modalities during the reconstruction process. However, the decoder is discarded in the downstream classification task, meaning any fusion information learned by the decoder would not be utilized. To avoid this waste of learning resources and ensure the encoder focuses on learning modality-specific robust representations, we design lightweight separate decoders for each modality to perform independent reconstruction. Each decoder consists of four layers, where bilinear interpolation is used for up-sampling, followed by a depthwise separable convolution \cite{8099678DSConv} to refine the feature maps. At each layer, the decoder takes as input both the output from the previous decoder layer and the corresponding encoder feature maps. To ensure compatibility, a convolutional layer $\delta_i^k(\cdot)$ projects the encoder features to the same spatial resolution and channel dimension as the decoder outputs, after which the two are fused via element-wise addition. The decoder block $\phi_i^k(\cdot)$ generates the output features as: 
\begin{equation}
\small
\quad D_{i}^{k}=\phi_i^{k}(\delta_i^{k}(\tilde E_i^{k})+D_{i+1}^{k}),
\end{equation}
where $i = 4,3,2,1$ indexes the layer, $\tilde E_i^k$ denotes the projected encoder feature maps from the $i$-th layer, and $D_{i+1}^{k}$ is the output from the previous decoder layer, with $D_{5}^{k}=\mathbf{0}$. The outputs of the decoders are $\{D_1^{f}, D_1^{o}, D_1^{v}\}$. 
Finally, we introduce a regression head to fulfill the image reconstruction task, where each modality-specific decoder output is passed through a separate convolutional layer to reconstruct the corresponding input image as follows: 
\begin{equation}
\small
\hat S^{k}_{i} = \rho^{k}(D_1^{k}),
\end{equation}
where $\rho^{k}(\cdot)$ denotes the convolutional reconstruction head for modality $k$.  

To train the model, we adopt the mean squared error (MSE) loss as the loss function, which can be calculated by: 
\begin{equation}
\small
\mathcal{L}_{MSE}=\frac{1}{N}\sum_{i=1}^{N}\sum_{k \in K}\sum_{p=1}^P (s_i^{k(p)}-\hat{s}_i^{k(p)})^2,
\end{equation}
where \(N\) is the batch size, \(K=\{f,o,v\}\) denotes the set of modalities, \(P\) is the number of masked pixels per image, \(s_i^{k(p)}\) denotes the pixel intensity at position \(p\) in the original image, and \(\hat{s}_i^{k(p)}\) represents the corresponding reconstructed pixel intensity. This reconstruction objective encourages the model to develop robust feature representations that preserve critical anatomical structures and pathological features despite significant input corruption.

\subsection{Fine-tuning for Multimodal Glaucoma Classification}
For the glaucoma classification downstream task, we perform fine-tuning on the pretrained model by discarding the decoder and retaining only the encoder for hierarchical feature extraction. Given the inputs \(\mathbf{X}^S\), the encoder outputs three modality-specific feature maps \(\{E_4^{f},E_4^{o},E_4^{v}\}\). For effective multimodal feature fusion, we utilize a simple but effective way \cite{10388423GeCoM-NET,9068414concat}, concatenating the feature vectors from three feature maps into an enhanced feature map \(E_4\). Subsequently, a global average pooling layer is employed to convert the 2D feature map into a compact 1D feature vector as follows: 
\begin{equation}
\small
v_\alpha^k=GAP(E_4^k),
\end{equation}
where, \(GAP\) denotes global average pooling layer, \(v_\alpha^k\) denotes pooled feature maps. Finally, we employ two fully-connected layers with ReLU as the classification head to classify the feature vector into specific categories, and we denote the output possibility as a positive case for each input multimodal sample \(\mathbf{X}\) as \(p\). We use Cross Entropy loss as the loss function. 
The Cross Entropy loss can be calculated by: 
\begin{equation}
\small
\mathcal{L}_{CE} = -\frac{1}{N}\sum_{i=1}^{N} \sum_{c=1}^{C} y_{ic}log(p_{ic}),
\end{equation}
where, \(N\) denotes the batch size, \(C\) denotes the number of classes, \(i\) indexes the \(i\)-th sample in the batch, \(c\) indexes the class label, \(y_{ic} \in \{0,1\}\) is the true binary indicator, and \(p_{ic} \in [0,1]\) is the predicted probability that sample i belongs to class \(c\), satisfying \(\sum_{c=1}^C p_{ic} =1\) for each sample \(i\).

\section{Experiments and Results\label{sec5}}

\subsection{Implementation details}

\subsubsection{Experiment Environment}

All experiments are conducted in a Kaggle cloud environment utilizing an NVIDIA P100 GPU with 16 GB memory. During model training, we employ the Adam optimizer. 
A learning rate of \(1\times10^{-5}\) is used for pre-training models, while fine-tuning for the classification task utilizes a learning rate of \(3\times10^{-6}\), yielding optimal performance across modalities. 
The dataset is split into training, validation, and test sets following a 6:2:2 ratio. 
Specifically, we adopt stratified random sampling at the eye level to ensure that the class distribution in each subset matches that of the entire dataset, maintaining class balance across all partitions. The subsets are available in our publicly accessible dataset to guarantee the reproducibility of the experiments. For each modality, input images are resized to a resolution of 224$\times$224, and we apply the canonical per-channel pixel normalization with the parameters: mean = [0.485, 0.456, 0.406] and standard deviation = [0.229, 0.224, 0.225]. Our MCGA is configured with 4 attention heads.

During Stage 1, models are pre-trained for 20 epochs with a masking ratio of 0.7 and a batch size of 8. During Stage 2, we employ a batch size of 16 and perform validation after each training epoch. Training continues until the validation loss has plateaued for 10 consecutive epochs, triggering early stopping.
To ensure statistical reliability, five independent training trials are conducted with random initialization, and the mean results across runs constitute the final performance scores.

\subsubsection{Data Augmentation}

To improve robustness and reduce overfitting, we apply data augmentation while training. SLO images undergo random resized crop, color jitter, and vertical flip; OCT images apply random color jitter; VF PD maps apply random vertical flip. Additionally, all three co-registered modalities share a synchronous horizontal flip to preserve anatomical consistency between left and right eyes. Flip operations are applied with 50\% probability, and color jitter varies brightness, contrast, and saturation within [0.9, 1.1].

\subsubsection{Evaluation metric}

To evaluate the performance of the proposed model, we employ accuracy, F1-score, the area under the ROC (receiver operating characteristic) curve (AUROC), and Quadratic Weighted Kappa (QWK) as the evaluation metrics, where larger values indicate better performance.
In addition, we use Expected Calibration Error \cite{ECE} (ECE) and Brier score to analyze model reliability
, where smaller values indicate better performance. ECE quantifies the discrepancy between predicted confidence and empirical accuracy across \(M\) probability bins as:
\begin{equation}
\small
\mathrm{ECE} = \sum_{m=1}^M \frac{|B_m|}{N} |\mathrm{acc}(B_m) - \mathrm{conf}(B_m) |,
\end{equation}
where \(B_m\) denotes the \(m\)-th bin containing samples with predicted probabilities in the interval \(I_m\), \(|B_m|\)
 is the sample count in bin \(B_m\), \(N\) is the total sample size, \(\mathrm{acc}(B_m)\) is the empirical accuracy in \(B_m\), and \(\mathrm{conf}(B_m)\) is the average predicted confidence in \(B_m\). The Brier score evaluates the mean squared deviation as:
\begin{equation}
\small
\mathrm{Brier} = \frac{1}{N}\sum_{i=1}^{N}\sum_{c=1}^C \left({p}_{ic} - y_{ic}\right)^2,
\end{equation}
where \(y_{ic} \in \{0,1\}\) is the true binary indicator, and \(p_{ic} \in [0,1]\) is the predicted probability that sample i belongs to class \(c\), satisfying \(\sum_{c=1}^C p_{ic} =1\) for each sample \(i\), thereby providing complementary measures of uncertainty calibration beyond traditional classification metrics.

\subsection{Comparison With SOTA Methods}

\begin{table*}[!t] 
\centering 
\renewcommand{\arraystretch}{1.} 
\renewcommand{\tabcolsep}{13.5pt} 
\caption{Comparison of the results obtained by HAMM and several representative methods on the GLEAM dataset. Statistical significance of accuracy differences relative to the baseline method is evaluated using two-tailed paired t-tests. \textit{(*: $p<0.05$; **: $p<0.01$, where $p$ denotes p-value.} TL: Transfer Learning, SCL: Supervised Contrastive Learning, SSL: Self-supervised Learning.}

\label{tab:Table 2} 
\begin{tabular}{r||ccccc|cc} 
\hline 
\Xhline{2.\arrayrulewidth} 
Method & Pre-train&Acc&F1 &AUROC&QWK&Param&FLOPs\\ 
\hline 
ResNet50\cite{2016ResNet} &-&76.75$\pm$1.47&66.84$\pm$2.60&89.95$\pm$0.27&85.88$\pm$0.76&\textbf{83.12}&\textbf{12.41}\\ 
\hline 
ResNet50\cite{2016ResNet}& TL& 77.67$\pm$0.86&70.19$\pm$0.93&92.14$\pm$1.81&87.00$\pm$0.55&\textbf{83.12}&\textbf{12.41}\\ 
ViT-S\cite{Vit}& TL& 77.75$\pm$1.52&69.62$\pm$3.79&91.79$\pm$0.48&88.03$\pm$0.64&65.21&12.75\\ 
ConvNeXt-T\cite{liu2022convnet} &TL& 79.00$\pm$0.76**&71.58$\pm$1.32&91.87$\pm$0.77&87.83$\pm$0.72&85.20&13.39\\ 
MHCA\cite{10030757MHCA} &TL& 78.16$\pm$0.63&69.97$\pm$3.20&92.28$\pm$0.27&87.14$\pm$0.75&248.29&17.63\\ 
DRIFA-Net\cite{10944200DRIFA-Net} &TL& 77.83$\pm$0.86*&69.70$\pm$1.96&92.42$\pm$0.10&86.75$\pm$0.57&931.31&88.48\\ 
\hline 
Corolla\cite{9761712Corolla} &SCL& 78.67$\pm$0.74**&72.87$\pm$1.21&92.39$\pm$0.55&88.50$\pm$0.32&83.90&\textbf{12.41}\\ 
ETSCL\cite{10.1007/978-3-031-73119-8_2ETSCL} &SCL& 79.08$\pm$0.80*&72.52$\pm$2.11&92.73$\pm$0.32&87.31$\pm$0.43&94.06&16.53\\ 
\hline 
MultiMAE\cite{MultiMAE}&SSL&78.00$\pm$0.18*&69.02$\pm$2.18&90.64$\pm$0.26&86.98$\pm$0.37&86.83&50.42\\ 
UrFound\cite{UrFound}&SSL&78.67$\pm$0.35*&70.67$\pm$1.46&92.49$\pm$0.44&87.86$\pm$0.48&65.21&12.75\\ 
\hline 
HAMM(ours)&SSL&\textbf{81.08}$\pm$0.63**&\textbf{75.90}$\pm$0.80&\textbf{93.03}$\pm$0.26&\textbf{90.07}$\pm$0.46&237.52&12.68\\ 
\hline 
\Xhline{2.\arrayrulewidth} 
\end{tabular} 
\end{table*}

To evaluate our method, we compare HAMM with several SOTA pre-trained approaches. We include commonly used backbones such as ResNet-50 \cite{2016ResNet}, ViT-S \cite{Vit}, and ConvNeXt-T \cite{liu2022convnet} with ImageNet-1K pre-trained weights, using the same approach as the baseline in \cite{WU2023102938GAMMA}. 
Attention-based methods (MHCA \cite{10030757MHCA}, DRIFA-Net \cite{10944200DRIFA-Net}), supervised contrastive learning methods (Corolla \cite{9761712Corolla}, ETSCL \cite{10.1007/978-3-031-73119-8_2ETSCL}), and self-supervised approaches (MultiMAE \cite{MultiMAE}, UrFound \cite{UrFound}) are also compared.

\begin{figure}
\centering
\includegraphics[width=1\linewidth]{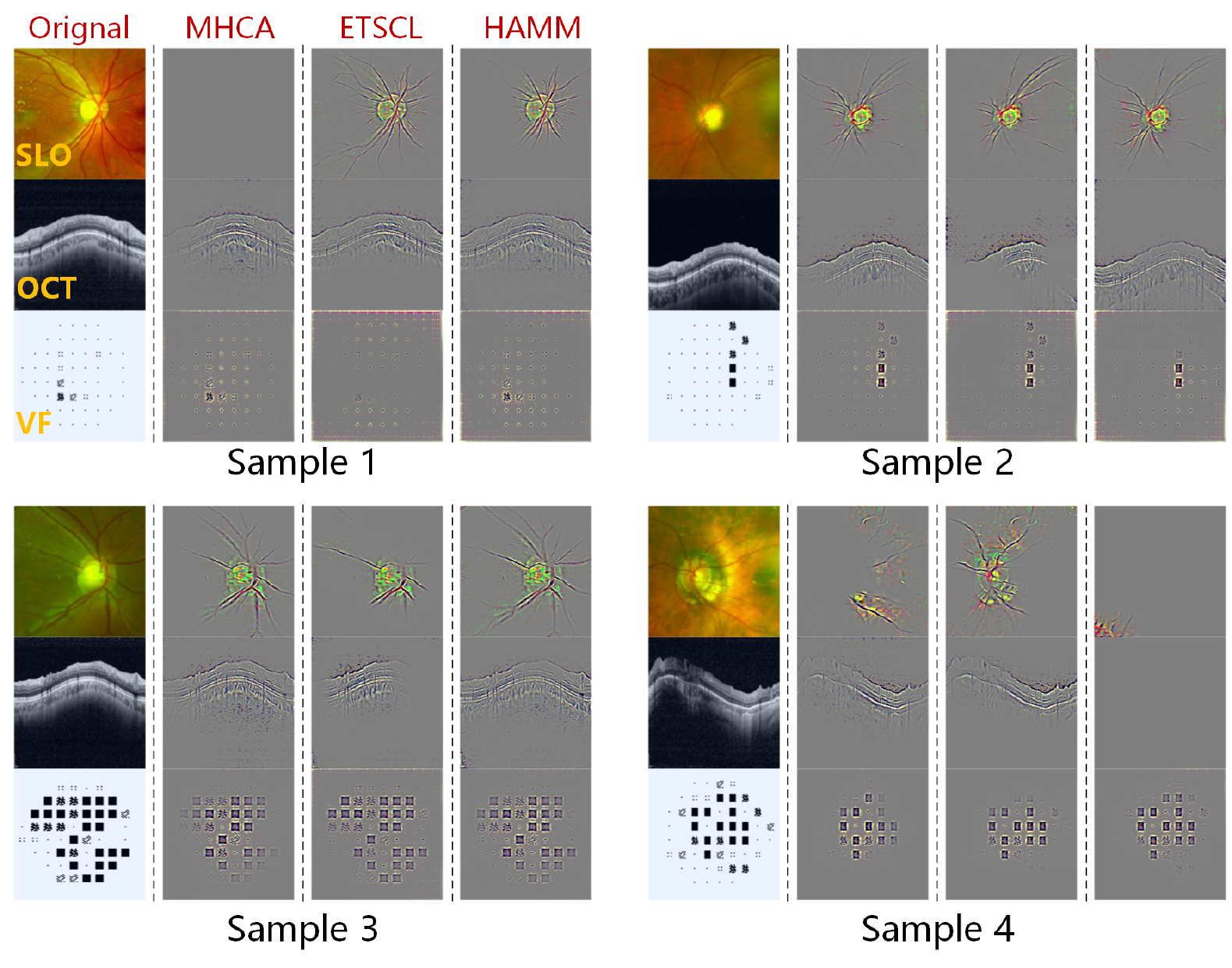}
\caption{Visualization of regions of interest identified by the model using Guided Grad-CAM}.
\label{fig 6}
\end{figure}
%

As shown in Tab. \ref{tab:Table 2}, 
HAMM achieves the best overall performance on the GLEAM dataset, with an accuracy of 81.08\%, F1-score of 75.90\%, AUROC of 93.03\%, and 90.07\% of QWK. For comparison, we next summarize the performance of representative methods. 
ResNet-50 performs better when using transfer learning method, with 0.92\% increase in accuracy, 3.35\% increase in F1-score, and 2.19\% increase in AUROC. Among transfer learning models, ConvNeXt-T outperforms ViT-S and ResNet-50, but still falls short of HAMM, with 2.08\% lower accuracy, 4.32\% lower F1-Score, and 1.16\% lower AUROC. Attention-based methods such as MHCA show improvement over ResNet-50, yet remain inferior to HAMM with 2.92\% lower accuracy, 5.93\% lower F1-score, and 0.75\% lower AUROC. SCL approaches, such as ETSCL, enhance classification by pre-training to better distinguish inter-class features, but still perform slightly below HAMM with 2.00\% lower accuracy, 3.38\% lower F1-score, and 0.3\% lower AUROC. SSL methods like MultiMAE achieves better performance than ViT through multimodal masked autoencoder, while UrFound utilizes knowledge-guided masked modeling to achieve competitive results. HAMM further employs a hierarchical attention module and light decoders to achieve superior integration of multimodal information, which yields the highest overall performance among all compared methods. 
In terms of computational efficiency, HAMM achieves superior FLOPs compared to most state-of-the-art methods, maintaining strong inference efficiency. While its parameter count is slightly higher than some lightweight baselines, it remains more parameter-efficient than other attention-based multimodal methods, such as MHCA and DRIFA-Net. To further optimize the model’s deployment potential, future work will explore more lightweight neural network architectures and knowledge distillation techniques. We will build on advances in compact multimodal fusion networks\cite{SHI2026113083} and medical image model distillation\cite{knowledge_distillation} in the future to reduce parameter overhead without compromising diagnostic performance.

Furthermore, we utilize guided gradient-weighted class activation maps\cite{Grad-CAM} (Guided Grad-CAM) to show the focus areas of the model for the provided prediction in the input images. 
We choose three models (MHCA, ETSCL, and HAMM) and visualize several test samples from GLEAM to examine model attention as shown in Fig. \ref{fig 6}. In the first sample (non-glaucoma), HAMM leverages ROI regions across all three modalities, while MHCA and ETSCL neglect SLO and VF images, respectively. In the second and third samples, both MHCA and HAMM integrate lesion areas from VF PD, fundus, and OCT, outperforming ETSCL. For the last sample, HAMM primarily focuses on VF, reflecting the pronounced VF deviations that allow accurate classification. In comparison, while MHCA and ETSCL also exhibit a tendency to downweight SLO and OCT modalities and shift their focus toward VF, they suffer from incomplete attenuation of these two modalities. This distinction underscores the critical role of the MCGA module in HAMM for effective modality weight calculation.

\subsection{Performance Analysis Across Modalities and Missing Modality}

\subsubsection{Performance Analysis Across Modalities}
To comprehensively evaluate the impact of different modalities on classification accuracy, we conduct experiments with various modality combinations, as shown in Tab. \ref{tab:Table 1}. Among single modalities, VF provides the strongest discriminative ability, while dual-modal settings further enhance performance, particularly when VF is included. The tri-modal fusion achieves the highest performance and more balanced accuracy across classes, confirming the complementary benefits of integrating multimodality.  It is worth noting that classification performance varies across severity classes even with balanced samples. Advanced glaucoma cases are classified more accurately than early and intermediate cases. This is because early and intermediate glaucoma involve subtle pathological changes with narrow clinical definition intervals, while advanced glaucoma exhibits obvious and distinct structural and functional abnormalities that are easier for the model to recognize.

\begin{table*}[!t]
    \centering
  \renewcommand{\arraystretch}{1.}
  \renewcommand{\tabcolsep}{13pt}
   \caption{Experimental results of different modalities. Statistical significance of accuracy differences relative to the tri-modal method is evaluated using two-tailed paired t-tests. \textit{(Acc-Cn: accuracy for class n.)}}
   \label{tab:Table 1}
  \begin{tabular}{r||cccc|cccc}
    \hline
    \Xhline{2.\arrayrulewidth}
Modality& Acc&F1 &AUROC&QWK&Acc-C1&Acc-C2&Acc-C3&Acc-C4\\\hline
 SLO& 60.25**&37.25 &74.72&56.79&94.83&3.00&2.00&72.00\\
 OCT& 61.75**&42.39 &76.70&62.84&93.33&8.00&10.50&72.00\\
  VF&74.25**&59.85 &90.42&83.03&\textbf{98.00}&6.00&57.50&88.00\\
  \hline
 SLO+OCT& 64.42**&46.47 &67.22&79.72&94.00&11.00&13.50&80.00\\
 SLO+VF& 77.67*&68.36 &91.87&86.84&96.33&26.00&61.50&89.50\\
 OCT+VF& 77.08*&67.38 &92.24&86.25&96.50&22.50&61.00&89.50\\
 \hline
 SLO+OCT+VF&\textbf{81.08}&\textbf{75.90}&\textbf{93.03}&\textbf{90.07}&93.67&\textbf{51.50}&\textbf{63.00}&\textbf{91.00}\\
    \hline
    \Xhline{2.\arrayrulewidth}
    \end{tabular}
  \end{table*}

\begin{figure*}
\centering
\includegraphics[width=0.95\linewidth]{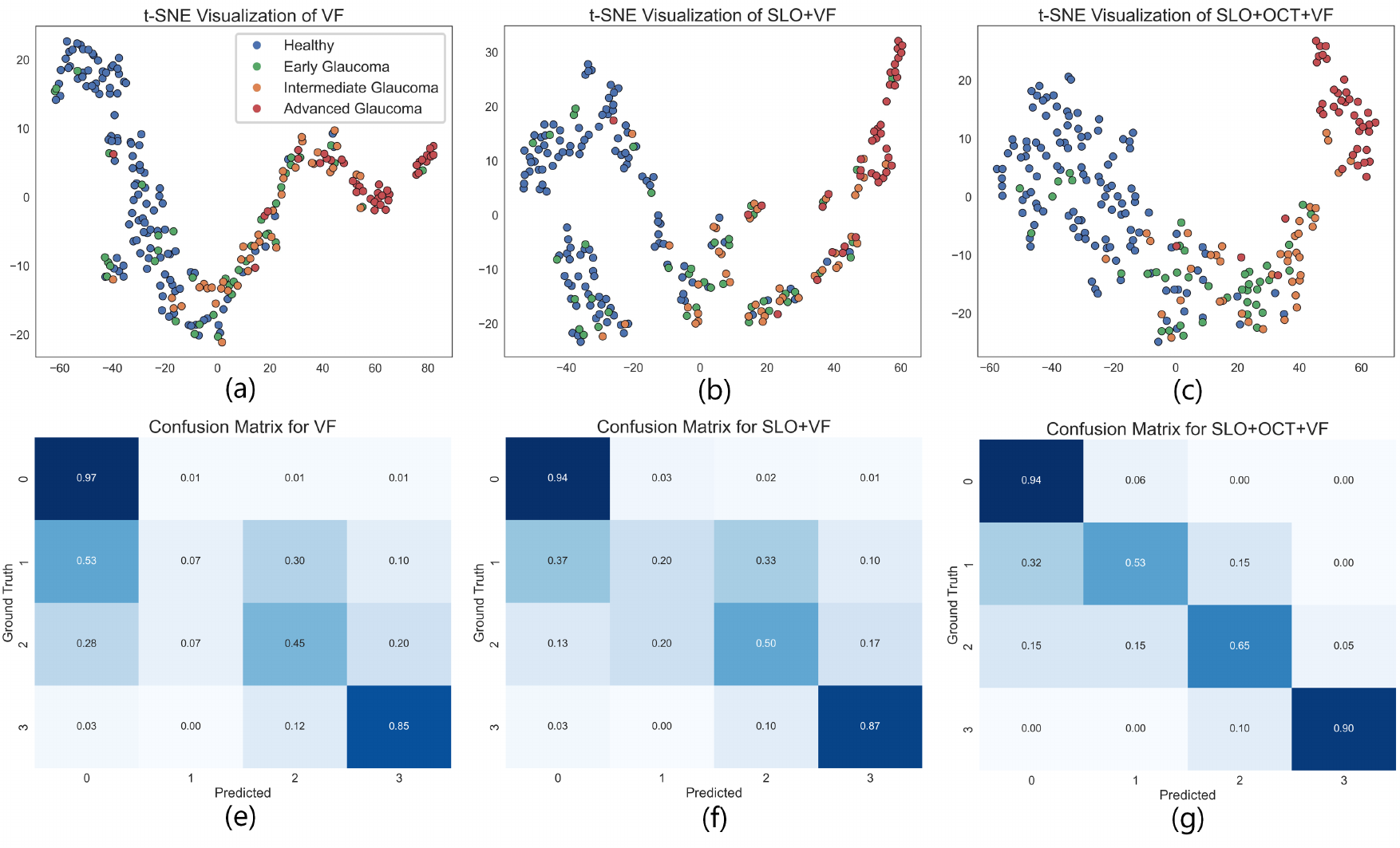}
\caption{Comprehensive evaluation of model feature representations and classification performance under three input settings: (a), (b), and (c) are t-SNE visualizations of feature embeddings corresponding to the input settings of VF, SLO+VF, and SLO+OCT+VF respectively; (e), (f), and (g) are confusion matrices corresponding to the identical set of input combinations.}
	\label{fig 7}
\end{figure*}

To further investigate the diagnostic value of multimodal fusion in glaucoma classification, we employ t-SNE 
dimensionality reduction to visualize feature representations extracted from optimally performing models under different modality combinations (VF, SLO+VF, and SLO+OCT+VF). As shown in Fig. \ref{fig 7} (a), the VF-only approach exhibits minimal inter-class separation between adjacent disease stages. Notably, substantial overlap persists between intermediate and advanced stage samples. In Fig. \ref{fig 7} (b), the dual-modality approach demonstrates a significant enhancement of class separation. The margin between early and intermediate stages widens considerably, and overlap between intermediate and advanced groups is markedly reduced. As shown in Fig. \ref{fig 7} (c), the tri-modality approach demonstrates optimal separation performance with maximized inter-class margins across all disease stages, and refines discrimination between early and intermediate stage cohorts. Complementing the feature visualization results, the confusion matrices in Fig. \ref{fig 7} (e), (f), and (g) further quantify the classification robustness of the three modality combinations in clinical diagnosis scenarios. Specifically, the VF-only model in Fig. \ref{fig 7} (e) nearly loses the capability of early-stage glaucoma diagnosis, indicating its limited practical utility for timely intervention. For the dual-modality model depicted in Fig. \ref{fig 7} (f), the diagnostic performance is improved compared with the VF-only setting, yet non-negligible misclassification phenomena remain: NG cases are incorrectly categorized as AdG, and AdG cases are misidentified as NG. By contrast, the tri-modality model in Fig. \ref{fig 7} (g) achieves the most substantial improvement in diagnostic accuracy; the aforementioned cross-misclassification issues between NG and AdG are completely eliminated, which verifies the advantages of integrating multimodality.

\subsubsection{Performance Under Missing Modality Scenarios}

In practice, not all patients undergo complete examinations, and certain modalities may be unavailable due to factors such as equipment limitations or patient non-compliance. To simulate these realistic scenarios, we design an experiment where missing modalities are introduced during both training and evaluation.

During training, for each sample of each training epoch, we randomly remove one modality with a probability of 25\%, two modalities with a probability of 25\%, and keep all modalities intact with a probability of 50\%. As a result, the same sample may have different missing modalities across epochs, helping expose the model to a variety of missing-modality scenarios. For the validation and test sets, we retain the original full-modality samples, then add new samples by modifying 50\% of the original samples to have one modality missing, and 50\% to have two modalities missing. In these modified sets, the distribution of missing modalities is consistent, with equal numbers of samples having each modality missing. This results in the test set having 480 samples (240 full-modality, 120 with one modality missing, and 120 with two modalities missing), and similarly, the validation set also grows to 480 samples after applying the missing-modality configurations. The subsets of missing modalities used for validation and testing are available in our publicly accessible dataset to guarantee the reproducibility of the experiments.

Tab. \ref{tab:missing_modality_results} shows the performance of different models under missing-modality conditions. As expected, the model’s accuracy declines when one or more modalities are missing. However, even when faced with missing modalities, HAMM achieves the best performance across all evaluation metrics. These results demonstrate that HAMM maintains strong robustness under incomplete input conditions, highlighting the effectiveness of the proposed multimodal fusion strategy in handling missing modalities.

\begin{table}[!t]
\centering
\renewcommand{\arraystretch}{1.3}
\renewcommand{\tabcolsep}{10pt}
\begingroup

\caption{Model performance under missing-modality scenarios.}
\label{tab:missing_modality_results}
\begin{tabular}{l||cccc}
\hline
\Xhline{2.\arrayrulewidth}
Modality & Acc & F1 & AUROC & QWK\\
\hline
Corolla & 69.87 & 60.42 & 84.65 & 73.15\\
ETSCL & 70.35 & 61.76 & 84.89 & 73.22\\
MultiMAE & 71.62 & 64.15 &85.76  & 75.25\\
URFound & 72.48 & 64.89 & 86.83 & 76.28\\
\hline
HAMM(Ours) & \textbf{74.59} & \textbf{66.73} & \textbf{88.92} & \textbf{77.05}\\
\hline
\Xhline{2.\arrayrulewidth}

\end{tabular}
\endgroup
\end{table}

\subsection{Reliability Analysis}
\begin{figure}
\centering
\includegraphics[width=1\linewidth]{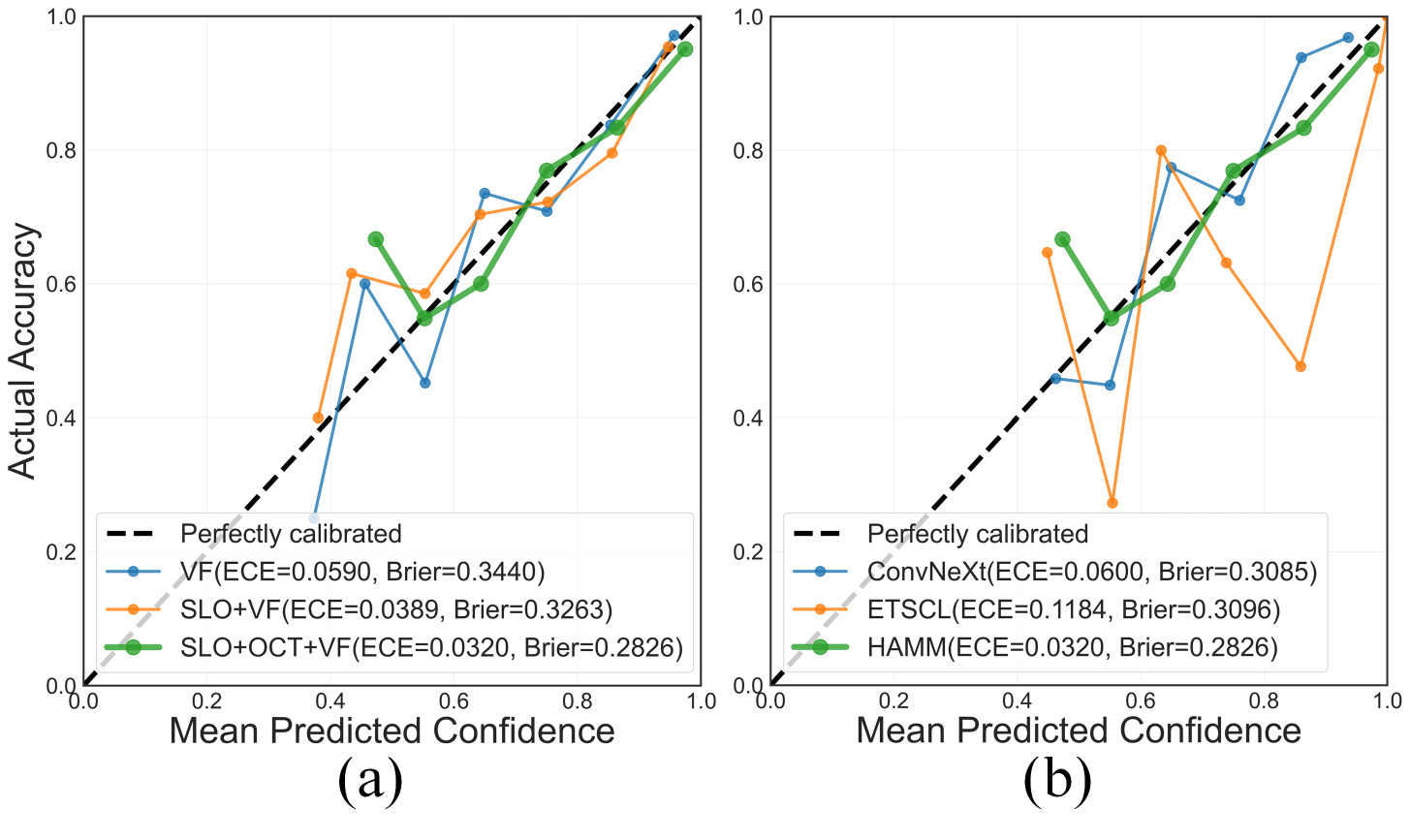}
    \caption{Reliability diagram of models, including: (a) reliability diagram for different modalities VF, SLO+VF, and SLO+OCT+VF, and (b) reliability diagram for SOTA methods ConvNeXt, ETSCL, and HAMM.}
	\label{fig 8}
\end{figure}

We further evaluate model calibration using reliability diagrams. As shown in Fig.~\ref{fig 8}(a), tri-modal fusion consistently achieves better calibration than single- and dual-modal settings, showing the most reliable confidence estimates. Fig.~\ref{fig 8}(b) compares the best three methods (ConvNeXt, ETSCL, and HAMM), where ETSCL exhibits poor calibration despite high accuracy, while ConvNeXt trades slight reliability for performance, and HAMM achieves both high accuracy and well-calibrated predictions. These results demonstrate that accuracy and reliability do not always coincide. However, our proposed HAMM effectively enhances both through pre-training and attention mechanisms. Notably, unlike single-modality and dual-modality, the tri-modality reliability curve lacked segments near the 0.4 confidence level, which indicates that the model assigns consistently higher confidence scores, consequently resulting in no samples falling within this range. These results demonstrate that multimodal fusion not only enhances accuracy but also improves confidence calibration, showing better reliability.

\subsection{Ablation Analysis}
\subsubsection{Component Ablation}

Ablation studies are conducted to disentangle the contributions of the MCGA module and the MAE mechanism. As shown in Tab. \ref{tab:Table 3}, both components independently enhance model performance across multiple metrics. MCGA shows relatively stronger improvements in AUROC, indicating its role in refining discriminative feature representations and reducing irrelevant activations. Our MAE provides more noticeable gains in accuracy and F1-Score, suggesting improved robustness and generalization through representation learning. When integrated, the two modules complement each other and deliver the best overall performance, highlighting their synergistic effect. Notably, our MCGA module alone outperforms other attention-based methods (MHCA and DRIFA-Net), and our MAE mechanism alone also achieves better performance than other MAE-based baselines, further verifying the superiority and rationality of each component's design.

\begin{table}[!t]
    \centering
  \renewcommand{\arraystretch}{1.}
  \renewcommand{\tabcolsep}{6pt}
   \caption{The ablation study results of HAMM.}
   \label{tab:Table 3}
  \begin{tabular}{cc||ccccc}
    \hline
    \Xhline{2.\arrayrulewidth}
MCGA&Pre-train&  Acc& F1&AUROC&QWK\\\hline
  && 77.67 & 70.19 & 92.14&87.00\\
    \Checkmark & & 79.17 & 71.93 & 92.89&89.52\\
  & \Checkmark & 79.67  & 73.68 &92.83&89.57\\
   \Checkmark&\Checkmark& \textbf{81.08}&\textbf{75.90}&\textbf{93.03}&\textbf{90.07}\\
    \hline
    \Xhline{2.\arrayrulewidth}
    \end{tabular}
  \end{table}

\subsubsection{Effect of MCGA }

\begin{table*}[!t]
    \centering
    \renewcommand{\arraystretch}{1.5}
    \renewcommand{\tabcolsep}{5pt}
    \caption{Ablation study results showing the contribution of different components in the MCGA module.}
    \label{tab:ablation_study}
    \begingroup
    
    \begin{tabular}{c|c|c||cccc|cc}
        \hline
        \Xhline{2.\arrayrulewidth}
        Pooling Scheme & Attention Scheme & Fusion Scheme & Acc&F1&AUROC&QWK&Param&FLOPs\\
        \hline
 -& CBAM\cite{CBAM} & Late&78.00& 70.89 & 92.14 & 87.93 &125.59&14.27 
\\
 -& CBAM\cite{CBAM} & Hierarchical&78.62 & 71.46&92.51& 88.42&139.53&19.83 \\
        \hline
        GAP only & Gating only & Hierarchical&77.75& 70.61 & 92.49 & 87.02 & 131.68&\textbf{12.49} \\
 GAP only & Graph only& Hierarchical&78.59& 71.15 & 92.41 & 87.92& 166.70&12.60 \\
        GAP only & Full Attention& Hierarchical& 78.75 & 71.82&92.66& 88.76 &233.55 &12.66 \\
        \hline
 Full Pooling& Full Attention& Late&78.50& 71.25 & 92.36 & 88.30 & \textbf{118.80}&12.46\\
        Full Pooling& Full Attention& Hierarchical& \textbf{79.17} & \textbf{71.93} &\textbf{92.89} &\textbf{89.52}&237.52&12.68\\
        \hline
        \Xhline{2.\arrayrulewidth}
    \end{tabular}
    \endgroup
\end{table*}

To further investigate the contribution of each component in the MCGA module, we evaluate its individual components in detail. For the pooling scheme, we compare our proposed approach with the commonly used GAP. For the attention scheme, in addition to ablating the two key components of MCGA, we also compare our attention mechanism with CBAM\cite{CBAM}, a widely adopted attention mechanism that applies both channel and spatial attention to refine feature representations. In our multimodal setup, we first concatenate the different modalities along the channel dimension, then pass the concatenated features through the CBAM attention mechanism. Finally, a convolutional projection layer maps the attention-enhanced features back to the individual modalities. 
Furthermore, we examine two fusion strategies: late fusion, which applies fusion attention after encoding all modalities, and hierarchical fusion, which performs fusion at each convolutional layer, allowing for more localized and progressive feature aggregation. 
As shown in Tab. \ref{tab:ablation_study}, the results of the ablation study confirm the effectiveness of the individual components in the MCGA module. Notably, when applying hierarchical fusion, while the parameter count increases significantly due to the additional attention mechanisms at each layer, the increase in FLOPs is minimal. This shows that our method achieves improved performance without a substantial computational cost. The combination of the full MCGA module in hierarchical fusion yields the highest performance across all metrics, further emphasizing its efficacy in enhancing multimodal fusion and feature representation.

\begin{figure}
    \centering
    \includegraphics[width=0.95\linewidth]{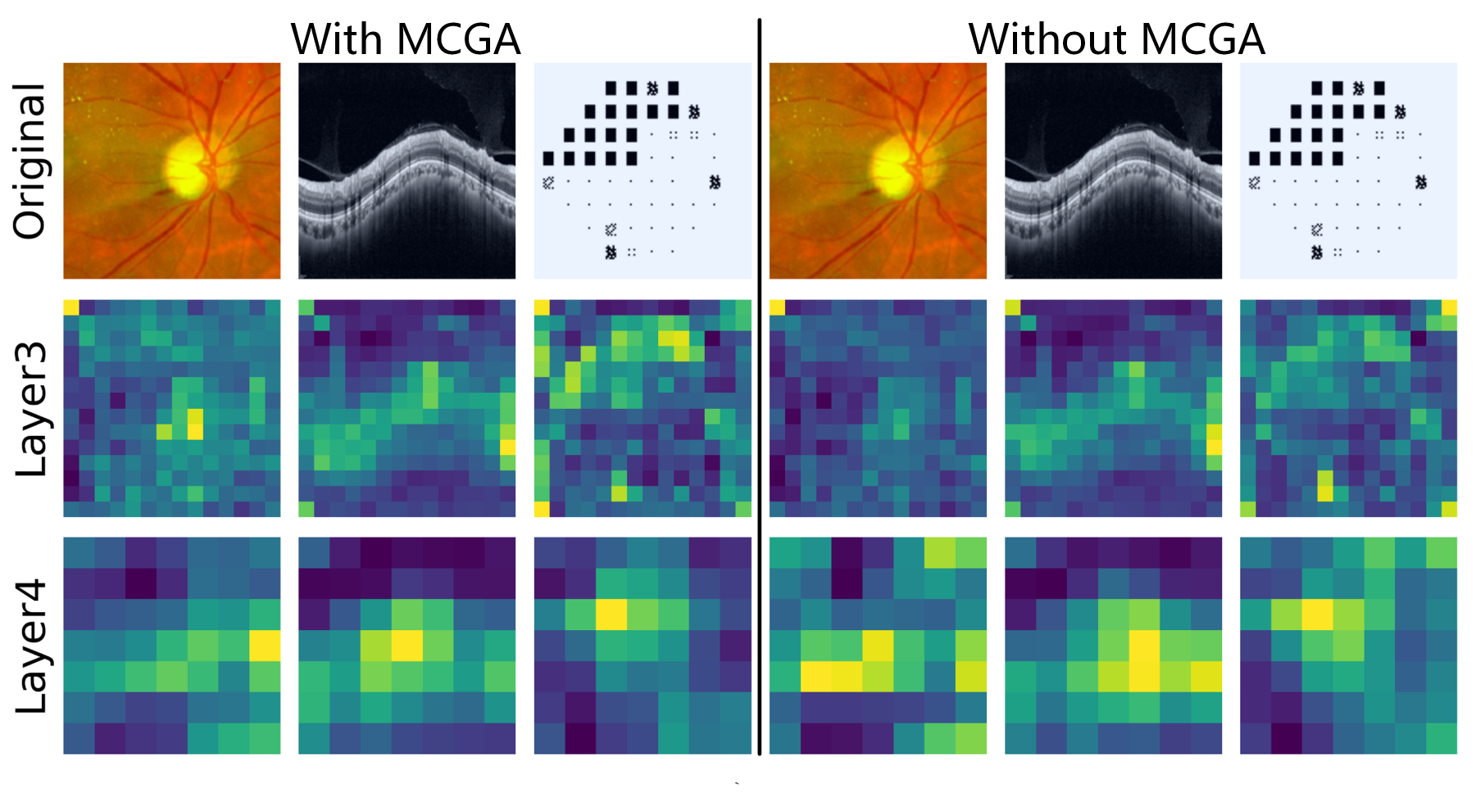}
    \vspace{-5pt}
    \caption{Visualization of averaged feature maps from the 3rd and 4th backbone layers with and without the MCGA module.}
    \label{fig 9}

\end{figure}

We also visualize the averaged feature maps from the 3rd and 4th layers of the backbone with and without MCGA. As shown in Fig. \ref{fig 9}, feature responses become more concentrated on informative multimodal regions when MCGA is applied, indicating that MCGA effectively enhances feature representation. 

\subsubsection{Masking Ratio Optimization Analysis}

\begin{figure}
    \centering
    \includegraphics[width=0.95\linewidth]{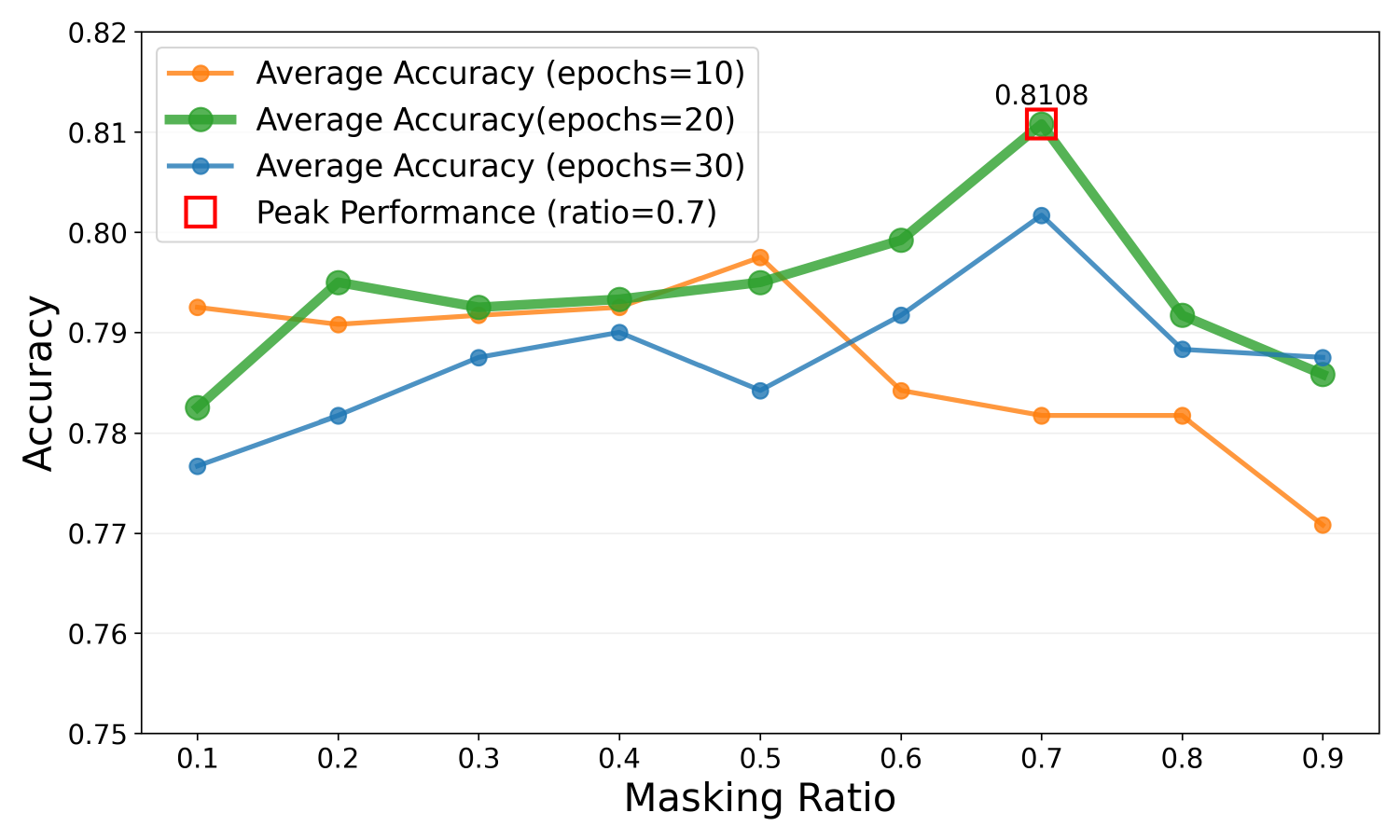}
    \vspace{-5pt}
    \caption{Impact of masking ratio on model accuracy, with the optimum at 0.7 when trained for 20 epochs. }
    \label{fig 10}

\end{figure}

To investigate the impact of masking ratio, we conduct experiments with ratios ranging from 0.1 to 0.9. As shown in Fig. \ref{fig 10}, the model accuracy increases steadily, peaking at 81.08\% with a masking ratio of 0.7 after 20 epochs of training, while further increase in the ratio leads to a noticeable decline. These results suggest that a moderate masking ratio, coupled with sufficient training, encourages the model to learn more discriminative features. Therefore, a masking ratio of 0.7 is identified as the optimal parameter configuration under the 20-epoch training setting for this task. 

\subsection{External validation on the GAMMA dataset}

We add an external validation experiment using the GAMMA dataset\cite{WU2023102938GAMMA}. GAMMA is a multimodal glaucoma classification dataset comprising paired 2D CFP images and 3D OCT volumes. As the official evaluation server of GAMMA reports only quadratically weighted Cohen’s Kappa, we follow this standard metric for all comparisons.

Since HAMM is designed to process multiple 2D image modalities, we convert the 3D OCT volumes into 2D RNFL thickness maps using the same preprocessing strategy as COROLLA\cite{9761712Corolla}. We apply the same data augmentation operations as COROLLA during training to maintain consistency with prior work on this benchmark.

The results on GAMMA are summarized in Tab. \ref{tab:gamma_method_compare}, where HAMM is compared with several recent SOTA methods. Given that many comparative methods reported performance using ensemble strategies, we not only provide the standalone performance of our single model but also supplement it with an ensemble result for fair benchmarking. For the ensemble implementation, we select the three models with the highest kappa values and compute the mean of their predictions. Furthermore, ETSCL\cite{10.1007/978-3-031-73119-8_2ETSCL} employs an additional modality branch, while Mstnet\cite{Wang_2023Mstnet} utilizes additional ophthalmic data for training. These two key characteristics have been explicitly indicated in the table. The results demonstrate that HAMM maintains competitive performance on this external dataset, suggesting robustness and generalizability across datasets with different acquisition protocols and modalities.

\begin{table}[!t]
    \centering
    \renewcommand{\arraystretch}{1.2} 
    \renewcommand{\tabcolsep}{4pt} 
    \caption{Comparison of Quadratic Weighted Kappa Results Among HAMM and SOTA Methods on the GAMMA dataset.\textit{(Add Info.: Use additional ophthalmic data for training, Add Mod.: Use additional modality)}}
    
    \label{tab:gamma_method_compare} 
    \begin{tabular}{l||c|c|c|c}
    \hline
        \Xhline{2.\arrayrulewidth}
        Method &Add Info.& Add Mod. & Ensemble & QWK \\
        \hline
        SmartDSP\cite{WU2023102938GAMMA} & \xmark& \xmark& \Checkmark & 85.49 \\
        COROLLA\cite{9761712Corolla} & \xmark& \xmark & \Checkmark & 85.50 \\
        ETSCL\cite{10.1007/978-3-031-73119-8_2ETSCL} & \xmark& \Checkmark & \xmark & 88.44 \\
        Mstnet (ResNet50)\cite{Wang_2023Mstnet}& \Checkmark&\xmark&\xmark&86.00\\
        GeCoM-Net\cite{10388423GeCoM-NET} & \xmark& \xmark& \xmark & 86.00 \\
        GeCoM-Net\cite{10388423GeCoM-NET} & \xmark& \xmark & \Checkmark & 88.10\\
        \hline
        HAMM(Ours) & \xmark& \xmark & \xmark & 87.59 \\
        HAMM(Ours) & \xmark& \xmark & \Checkmark& \textbf{89.35} \\
        \hline
        \Xhline{2.\arrayrulewidth}
    \end{tabular}
\end{table}

\section{Discussion\label{sec6}}
HAMM demonstrates significant advantages in glaucoma classification by effectively integrating multiple imaging modalities. By leveraging hierarchical attention mechanisms, HAMM not only learns robust multimodal feature representations but also emulates clinical reasoning by assigning adaptive weights to each modality. This enables the model to capture the interplay between structural and functional data, improving the overall classification performance. The results show that HAMM outperforms single-modality and traditional dual-modality approaches, making it a promising tool for accurate glaucoma diagnosis and staging.

While our results are promising, several data-related limitations merit discussion to guide future work. VF-derived labels may not always be available in clinical practice due to patient non-compliance or equipment constraints. Therefore, we plan to explore the model’s performance when VF data is excluded or when the model is trained with stratified MD bins to research its applicability in such scenarios. The current dataset is collected from a single clinical center using specific imaging devices, which may introduce biases that limit generalizability across diverse clinical environments; expanding to multi-center data and exploring domain adaptation techniques will help improve the model’s robustness in real-world settings. Glaucoma subtype heterogeneity also remains unaddressed in the current dataset, as different subtypes such as primary open-angle glaucoma and normal-tension glaucoma exhibit distinct pathophysiological features and damage patterns. Adding explicit subtype labels and exploring subtype-aware training strategies will ensure the model captures true stage-related patterns rather than subtype-specific variations, further enhancing its clinical utility. Another important limitation is the dataset splitting strategy. In this study, data are split at the eye level rather than the patient level, meaning that images from the left and right eyes of the same patient may be allocated to different subsets. Given that fellow eyes often share similar anatomical structures and disease characteristics, this may introduce inter-sample correlation across data partitions and lead to slightly optimistic performance estimates. Future work will adopt patient-wise splitting to provide a more rigorous evaluation.

Another area for improvement is the pretraining objective. Currently, we use independent modality reconstruction to guide encoder learning, which ensures robust unimodal feature extraction but lacks explicit enforcement of cross-modal structure–function consistency. Future work will integrate a cross-modal alignment loss to strengthen the model’s understanding of the interplay between structural and functional modalities. This alignment loss will mimic the clinical practice of cross-referencing structural and functional data, further enhancing the model’s clinical relevance and diagnostic reliability.

\section{CONCLUSION\label{sec7}}
In this paper, we introduced GLEAM, the first publicly available tri-modal glaucoma dataset, and proposed HAMM, a novel multimodal fusion network designed to effectively capture inter-modal dependencies and enhance feature robustness through the MCGA module and masked modeling. Comprehensive evaluations demonstrate that tri-modal fusion consistently yields superior diagnostic accuracy and reliability compared with single-modal or dual-modal approaches. HAMM also surpasses a range of SOTA methods, while maintaining well-calibrated predictions.

\par  Beyond improving automated glaucoma detection, our work provides the community with a tri-modal dataset and open-source implementation, supporting exploration of intrinsic relationships among modalities to uncover underlying physiological mechanisms of glaucoma.  By facilitating a deeper understanding of disease progression, this dataset contributes to early diagnosis and ultimately promotes public eye health. Future research may explore domain adaptation, semi-supervised learning, longitudinal analysis, and missing modality problem to further enhance diagnostic performance and generalizability.

\bibliographystyle{IEEEtran}
\bibliography{IEEEabrv,sample}
\balance
\end{document}